\documentstyle[pre,aps,epsf,multicol]{revtex}

\begin{document}

\title{ Temperature Dependence of Facet Ridges in Crystal Surfaces.}
\author{ Douglas Davidson and Marcel den Nijs}
\address{ Department of Physics, University of Washington, P.O. Box 351560, \\
Seattle, Washington 98195-1560}
\maketitle

\begin{abstract}

The equilibrium crystal shape of a body-centered solid-on-solid (BCSOS) model 
on a honeycomb lattice is studied numerically.
We focus on the facet ridge endpoints (FRE).
These points are equivalent to one dimensional KPZ-type
growth in the exactly soluble square lattice BCSOS model.
In our more general context the transfer matrix is not stochastic at 
the FRE points, and a more complex structure develops. 
We observe ridge
lines sticking into the rough phase where the
surface orientation jumps inside the rounded
part of the crystal.  Moreover, the rough-to-faceted edges
become first-order with a jump in surface orientation, between
the FRE point and
Pokrovsky-Talapov (PT) type critical endpoints.
The latter display anisotropic scaling with exponent $z=3$ 
instead of familiar PT value $z=2$.

PACS number(s):  68.35 Bs, 64.60 Fr, 64.60 Ht, 68.35 Rh
\end{abstract}

\begin{multicols}{2}
\narrowtext

\section{Introduction}\label{sec:1}

Equilibrium crystal shapes typically consist of various flat facets
connected by  rounded surfaces.
The way this structure changes with temperature has been studied for
many years
~\cite{Wulff,Herring,Wortis,Rottman,Nolden,Grimb,JN-MdN,Buk-Shore,Buk-Shore2}.
Most aspects are well established. For example,
rounded parts are associated with crystal orientations where the 
surface is rough, and flat facets can disappear from the equilibrium shape
at the roughening temperature of those specific facets.
Roughening transitions belong to the 
Kosterlitz-Thouless (KT) universality class~\cite{KT}.
Many details follow from the scaling properties of the KT transition,
e.g., that the facet diameter vanishes exponentially with reduced temperature,
and that the curvature of the surface is universal just above the KT transition 
at that facet orientation.

Flat facets typically connect smoothly to rounded, rough surface areas.
The surface orientation in the rounded part varies continuously
and connects without a jump in angle onto the flat facet.
Such phase boundaries are described by 
Pokrovsky-Talapov (PT) transitions~\cite{Pok-Tal,MdN-IC,Fisher}.
The angle difference vanishes with a $\frac{3}{2}$ power.
The most salient feature of PT transitions is their anisotropic scaling.
The rounded surface on the rough side of the PT transition
can be interpreted as a stepped surface from the perspective of the flat facet. 
The characteristic lengths along, $\xi_\parallel$, 
and perpendicular to these steps, $\xi_\perp$,
scale as  $\xi_\parallel\sim \xi_\perp^z$, with $z=2$.

Two dimensional equilibrium statistical mechanical systems
are mathematically equivalent to 
one dimensional quantum systems at zero temperature. 
The transfer matrix formalism of the first 
is the path integral representation of the second.
PT transitions are equivalent to metal-insulator transitions
in one dimensional fermion systems. 
The steps represent the world lines of fermions, and the 
lattice direction parallel to the steps plays the role of time.
In the rough phase (the metal phase) the fermions behave relativistically,
i.e., time, $\xi_\parallel$, scales compared to space, $\xi_\perp$, 
with dynamic exponent $z=1$, $\xi_\parallel \sim \xi_\perp$.
At the PT metal-insulator transition the fermions behave non-relativistically,
with dynamic exponent $z=2$ instead of $z=1$~\cite{MdN-IC,Fisher}. 

The rough-to-facet interface can also be first-order,
with a discontinuity in the surface orientation angle
instead of the PT-type smooth connection.
The critical endpoint  between  such first-order and 
PT-type segments provides an example of a quantum field theory
with dynamic exponent $z=3$.
To our knowledge an explicit example of such a PT critical 
endpoint (PTE) has not been reported before. 
We will do so in this paper. 

Our current interest in these issues stems from the discovery that
the scaling properties at facet ridge endpoints (FRE) 
are related to KPZ-type growth~\cite{JN-MdN}.
FRE points are the endpoints of ridges where two facets connect directly.
The ridge splits into two PT or two first-order transitions with a
rounded, rough surface in between.
The exactly soluble body-centered solid-on-solid (BCSOS) model
on a square lattice (with only next nearest neighbour
interactions) contains such a FRE point ~\cite{Buk-Shore}.
Moreover, the transfer matrix of this model becomes stochastic at the FRE point
and maps exactly onto the master equation time evolution operator
for the so-called brick model in one dimension 
lower~\cite{JN-MdN,Dhar,Gwa-Spohn}.
The latter describes a growing 1D crystal.
The 1D growing interface is rough,
and its scaling properties  belong to the KPZ universality class.
The $\perp$-direction represents the spatial direction
and the $\parallel$-direction  represents time, just as in mapping of the
PT transtion to 1D fermions mentioned above.
The dynamic exponent is equal to $z=\frac{3}{2}$.
The characteristic time needed to reach the stationary growing state
diverges as $\xi_\parallel \sim N^z$, with $N$ the size of the 1D lattice.

\begin{figure}
\centerline{\epsfxsize=4cm \epsfbox{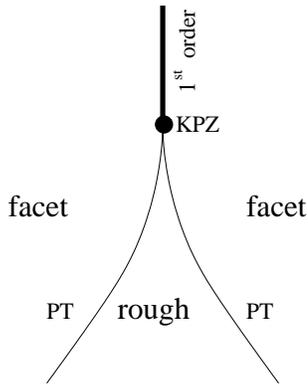}}
\caption{ Crystal structure with facet ridge endpoint as realized in the 
BCSOS model on a square lattice with only next nearest neighbour interactions.}
\label{6vertex}
\end{figure}

This translates into non-trivial scaling properties 
of KPZ-type FRE points~\cite{JN-MdN,Buk-Shore,Buk-Shore2}.
When approached from the rounded phase in the direction
parallel to the facet ridge, see Fig.~\ref{6vertex}, 
the surface orientation jumps. 
Moreover, at the rough side the characteristic length scale in the direction 
along the facet ridge, $\xi_\parallel$, 
scales as $\xi_\parallel\sim \xi_\perp^z$  with $z=\frac{3}{2}$.

This exact mathematical mapping between the scaling properties of 
equilibrium phase transitions to those of dynamic non-equilibrium transitions  
in one lower dimension is intriguing.
But it leaves us with an important question.
The above BCSOS model satisfies a special symmetry at its FRE point.
The transfer matrix is ``accidentally" stochastic.
It has  full rotational symmetry in the 
spin-$\frac{1}{2}$ operators (the step variables). 
This implies, e.g., a complete degeneracy of the free energy 
with respect to  all surface orientations.
Such a symmetry is too special.   
FRE points are a more general phenomenon, 
not limited to KPZ-type stochastic transfer matrices.
We need to investigate whether the KPZ-type scaling exponent
$z=\frac{3}{2}$ is  generic. 
Maybe stochasticity is restored at large length scales,
asymptotically close to the FRE point, 
with non-stochasticity  an irrelevant operator.
More likely, the FRE point changes its character.
In this paper we present such an investigation, 
a numerical study of a model with a complex crystal shape.

We define this model in section 2.
It originates from our earlier study of possible
spontaneous low-angle faceting in FCC(111)-type facets~\cite{DD-MdN}.
This model has the required property that its crystal shape is complex 
and that accidental symmetries are unlikely.
Section 3 contains an overview of the phase diagram and crystal shapes.
In section 4 we describe our numerical method.

The crystal structure is actually much richer than we anticipated.
Entire facets vanish as functions of the coupling constants.
FRE points become triple points.
We also find a more exotic feature, 
like a first-order (FOR) line inside the rough phase emerging 
from the FRE point (section 5).
At this line the surface orientation angle inside the rough phase jumps.
In their paper about the exact solution of the FRE point in the
square lattice BCSOS model, Bukman and Shore~\cite{Buk-Shore2}
speculate about the possible generalizations of Fig.~\ref{6vertex}.
The FOR line structure corresponds to their so-called
ridge scenario.  However, they doubted that it would be
realized.  In our model, FOR lines are clearly present
although they remain very short.

In a nearby section of the phase diagram, the FOR lines disappear.
The rough-to-faceted ridge becomes sharp with a jump in the surface
orientation angle.  These first-order lines terminate
further away from the FRE point 
in PTE points and the edge continues as a second order
PT line. 
In section 6 we study the scaling properties at these PTE points.
They display  anisotropic scaling with exponent $z=3$.
The conclusions are presented in section 7.

\section{The Model}\label{sec:2}

Our model emerges naturally at a coarse grained level in a study of
spontaneous low angle faceting in FCC(111) facets.
This will be described elsewhere~\cite{DD-MdN}.
For clarity it is more appropriate to introduce the model
from a different perspective.
Consider the (001) facet of an HCP-type crystal,
with  $ABABA$-type stacking of triangular slices.
The solid-on-solid (SOS) model description of such surfaces
leads to a BCSOS model on a honeycomb lattice.
The surface heights on the $A$-type triangular sublattice are even integers,
$h_A= 0,\pm 2, \pm 4, \dots$,
and those on the $B$-type triangular sublattice are odd integers,
$h_B= \pm 1,\pm 3,\pm 5, \dots$
The simplest  Hamiltonian for such a surface is
\begin{equation}\label{hamL}
H =  \frac{1}{4}L \sum_{(i,j)} (h_i-h_j)^2
\end{equation}
with $(i,j)$ the summation over next nearest neighbour columns.

We break the symmetry between the $A$- and $B$-type columns, such that
two distinct (001) facets can be cut from this crystal,
those with $A$'s or those with $B$'s on top.
$A$ and $B$ could be different type of molecules,
but that is not what we have in mind.
The interactions between the next nearest neighbour 
columns, $L_A$ and $L_B$, would then be different. 
Instead, assume that $A$ and $B$ are identical types of atoms,
but that for some reason the bulk structure is reconstructed
such that the  $A$'s move downward from the positions
exactly in-between the  $B$ layers.
In that case the bonds between the $A$ atoms with the $B$'s in the layer
beneath them are stronger than with the $B$'s in the layer above them.
We can model this by adding a nearest neighbour coupling
in the BCSOS model
\begin{equation}\label{hamK}
H = K \sum_{<i,j>} (h_i-h_j) + \frac{1}{4}L \sum_{(i,j)} (h_i-h_j)^2
\end{equation}
where, in the sum over nearest neighbors, $i$  ($j$) always refers
to $A$ ($B$) sublattice.

An alternative more generic scenario is the one realized in 
Naphthalene~\cite{Grimb} (for a different crystal symmetry however).
It involves only a surface reconstruction, no bulk reconstruction.
Consider a crystal where the $A$'s and $B's$ are identical molecules,
but with an orientation degree of freedom pointing in two different directions.
Inside the bulk these two orientations can be equivalent
by (a slide-mirror-plane type) symmetry, but the presence of
the surface typically breaks that equivalence.
The surface molecules rotate due to surface relaxation.
Typically they do so differently with the $A$'s or the $B$'s on top.
This breaks the $A$ $B$ symmetry  locally close to the surface
and creates an energy difference, $K\neq 0$.
The next nearest neighbour interactions remain virtually equal,
$L_A\simeq L_B=L$, if the orientational aspects of the  
interactions decay rapidly with distance.

\begin{figure}
\centerline{\epsfxsize=8cm \epsfbox{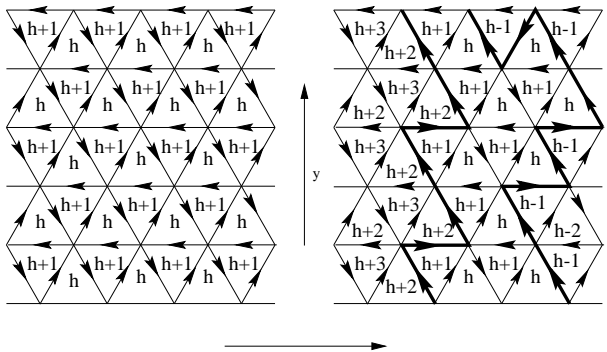}}
\caption{ Lattice of the triangular BCSOS model defined in eq.(3).}
\label{lattice}
\end{figure}

We introduce tilt energies $E_i$, $i=1,2,3$,
to study the crystal structure at and nearby 
the central (001) facet orientation.
The functional dependence of the free energy
with respect to these $E_i$ yields the
crystal shape by means of the Wulff construction~\cite{Wortis}.
\begin{eqnarray} \label{hamE}
{\cal H} = &  \frac{1}{4}L \sum_{(i,j)} (h_i-h_j)^2+ \nonumber\\
  &   (K+E_1) \sum_{<i,j>_1} (h_i-h_j) +  \nonumber\\
  &  (K+E_2) \sum_{<i,j>_2} (h_i-h_j) +  \nonumber\\
  &   (K+E_3) \sum_{<i,j>_3} (h_i-h_j) .
\end{eqnarray}
The summations run over nearest neighbour columns in the three
different directions, see Fig.~\ref{lattice}.
The three tilt fields $E_i$ are related as
\begin{eqnarray} \label{tilts}
E_1 & = & E_{x}  \nonumber\\
E_2 &= &  \frac{\sqrt{3}}{2} E_y - \frac{1}{2} E_{x}  \nonumber\\
E_3 &= & -\frac{\sqrt{3}}{2} E_y - \frac{1}{2} E_{x}
\end{eqnarray}
$E_{x}$ and $E_{y}$ are  conjugate to the tilt angles in the $x$- and $y$-directions,
i.e., parallel  and perpendicular to one of the three main axes of the
honeycomb lattice.

\begin{figure}
\centerline{\epsfxsize=8cm \epsfbox{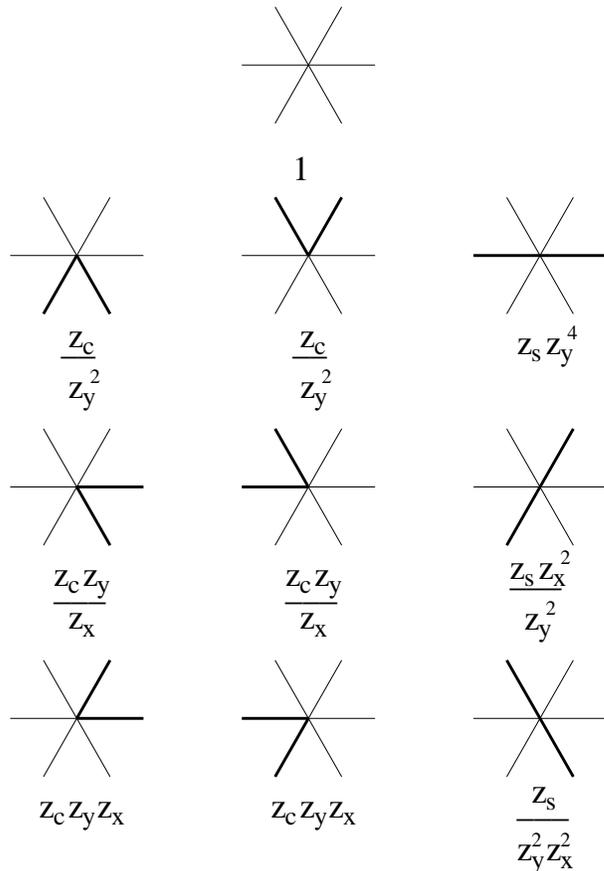}}
\caption{ All possible local configurations around each vertex
and their Boltzmann weights.}
\label{vertices}
\end{figure}

We simplify the model further by not allowing
step excitations to touch each other.
Fig.~\ref{vertices} shows all the remaining local configurations
in this 10-vertex model.
The arrows denote whether nearest 
neighbour $AB$ columns differ in height by $+1$ or $-1$.
The thin (fat)  lines denote that the  $K$ bond is
in the low (high) energy state.
The Boltzmann weight contributions are shown in Fig.~\ref{vertices}
below each vertex configuration, with 
$z_c=e^{-2K-2L}$,
$z_s=e^{-2K-4L}$,
$z_x=e^{-E_x/4}$, and
$z_y=e^{-\sqrt{3} E_{y}/4}$.
The coupling constants are made
dimensionless by absorbing the factor $k_BT$.

\section{The Phase Diagram}\label{sec:3}

\begin{figure}
\centerline{\epsfxsize=8cm \epsfbox{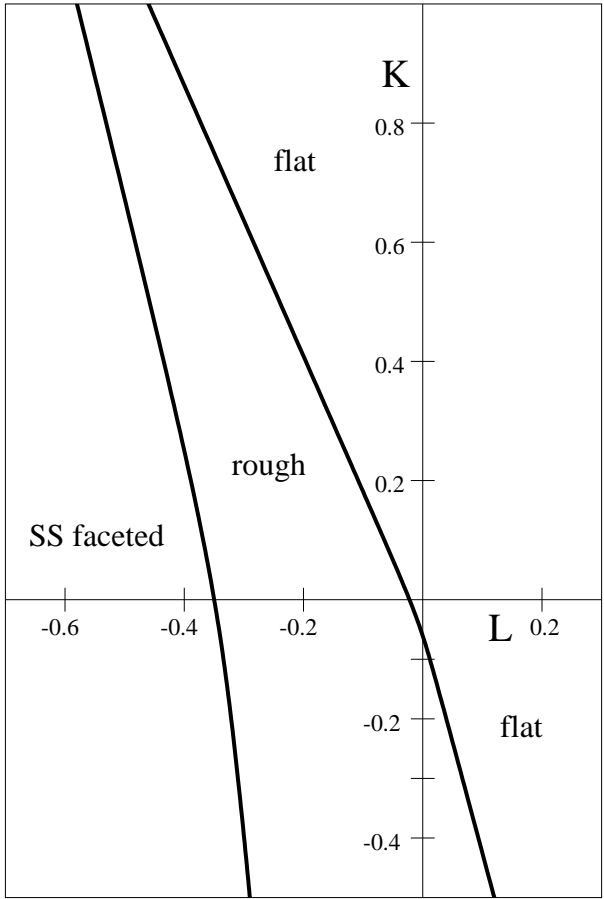}}
\caption{ Phase diagram of the triangular BCSOS model at zero tilt,
fields $E_x=E_y=0$.}
\label{phdgr}
\end{figure}

The phase diagram of our model at zero tilt fields $E_x=E_y=0$, 
Eq.(\ref{hamK}), is shown in  Fig.~\ref{phdgr}.
The (001) facet is stable at all temperatures inside the quadrant where 
$K$ and $L$ are both positive.
The step excitations do not create sufficient entropy to roughen this surface
even at $T=\infty$ (at point $K=L=0$).
The roughening line lies inside the other three quadrants.
We determine its location in the conventional manner for
transfer matrix finite size scaling calculations~\cite{MdN-87}.
The step free energy $\eta$ in  semi-infinite
strips of width $N$ is equal to the difference in free energy for
periodic and stepped boundary conditions in the finite direction.
The KT roughening line is obtained by extrapolating the lines where
$N\eta=\frac{1}{4}\pi$ 
for increasing strip widths. 

\begin{figure}
\centerline{\epsfxsize=8cm \epsfbox{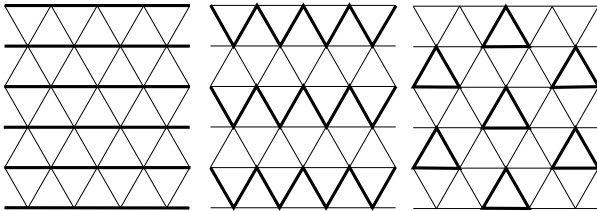}}
\caption{ Surface configurations in the:
(a) straight-step (SS) type facets,
(b) zig-zag (ZZ) type facets, and
(c) reconstructed triangles ground state.}
\label{facets}
\end{figure}

At low temperatures for $L<0$ the surface spontaneously facets into three
coexisting orientations. We refer to these  as straight-step (SS) facets, 
shown in Fig.~\ref{facets}(a)to distinguish them from the 
three zig-zag (ZZ) structures, shown in  Fig.~\ref{facets}(b), 
that appear elsewhere in the phase diagram at non-zero tilt fields $E_i$.
The transition from the rough phase into this phase with 
three coexisting SS facets is first-order. 
The surface orientation angle jumps. 
The location of this transition line is determined by the methods
described in section 4. Its location is not trivial,
unlike the square lattice BCSOS model~\cite{Nolden}
where it is simply the meeting point of several 
$E_i\neq 0$ PT transition lines at $E_i=0$.

An additional ground state appears 
in the quadrant with  $K<0$ and $L>0$,
the triangles-state shown in Fig.~\ref{facets}(c).
This state represents a highly degenerate surface reconstruction. 
Unfortunately  this ground state is unstable against thermal
fluctuations immediately at $T>0$ in this particular model. 
The surface reconstruction and associated preroughening phenomena
associated with this type of ground state could have been quite interesting. 

\begin{figure}
\centerline{\epsfxsize=8cm \epsfbox{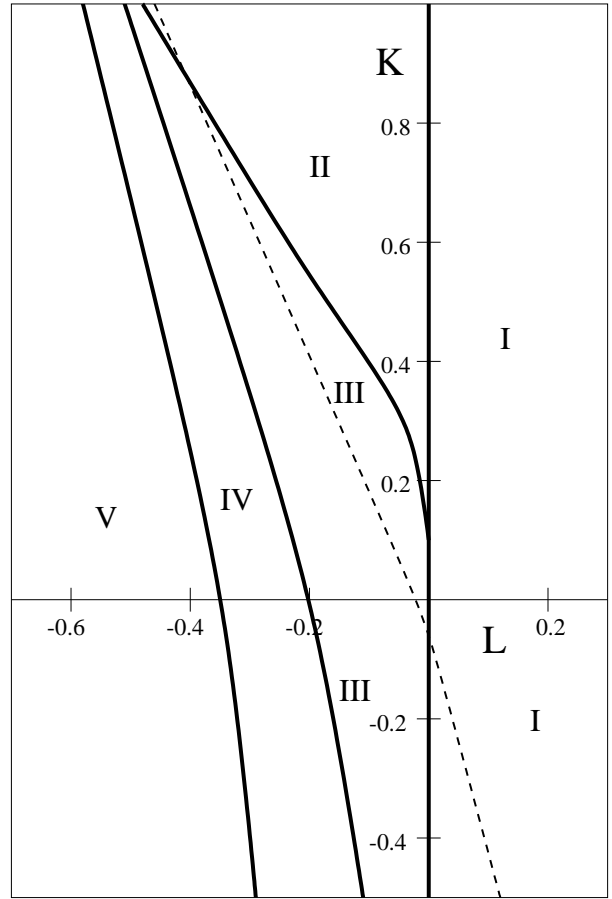}}
\caption{ Regions in the phase diagram where the various 
crystal shapes, {\bf I-V} as described in the text are realized.
The dashed line shows the location of the roughening line in the
zero field model.}
\label{phshapes}
\end{figure}

We determine the crystal shape at each value of $K$ and $L$  by
introducing the tilt fields $E_i$, see Eq.~(\ref{hamE}).
The technical details of these calculations are described in section 4.
In this section we summarize the crystal shape development with temperature.
We identified  at least 5 distinct crystal structures,
labeled {\bf I} through {\bf V} in Fig.~\ref{phshapes},
which shows at which values of $K$ and $L$ they are realized, 
as an overlay of the zero-tilt phase diagram  Fig.~\ref{phdgr}.

\begin{figure}
\centerline{\epsfxsize=4cm \epsfbox{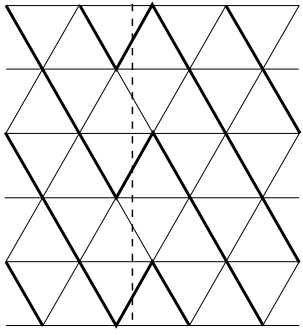}}
\caption{ SS facet with defect line (dashed).}
\label{defect}
\end{figure}

In region {\bf I}, located at $L>0$, the crystal structure consists of 
a non-tilted flat facet surrounded by a rough, rounded central region, 
and fully tilted facets of both types, SS and ZZ, at the edges
(each repeated three times).
This structure is shown in Fig.~\ref{shape1}.
All facets are separated from the rough region by PT lines.  
Their exact location  can be calculated easily for the SS  facets.  
Compare the free energy density of the faceted phase with that of a state 
with one line defect, e.g., as shown in Fig.~\ref{defect}.
The PT transition occurs when the free energy of the faceted phase becomes 
equal to the free energy with the  defect.
The calculation is straightforward, but leads to a somewhat complex formula.

The PT lines in region {\bf I} never touch or join.
There are no FRE points between the SS and ZZ facets.
The following analysis explains why.
Suppose facet ridges between the ZZ and SS facets exist.
Both faceted structures are frozen, without any fluctuations
(in our model) and therefore the free energy in both is equal to the energy,
$E_{ZZ}= 2K+2L-\frac{1}{2}E_x$  and 
$E_{SS}= 2K+4L-\frac{1}{2}E_x - \frac{1}{2} \sqrt{3} E_y$.
At a facet ridge these must be equal.
This happens at  $E_y = \frac{4}{\sqrt{3}}L$.  
However, the ZZ and SS facets roughen before this line is reached.
Consider the zero temperature limit of the $E_{ZZ}=E_{SS}$ line.
Line defects such as the one shown in Fig.~\ref{defect}
have the same energy as the two faceted configurations.
The facets are unstable with respect to these
defects, because at non-zero temperatures they gain entropy 
while the facets remain frozen.
Two  PT lines emerge from zero temperature and preempt the 
facet ridge. 

As $L$ decreases, the ZZ facets shrink and disappear at $L=0$.  
At $L\leq 0$, the only remaining fully tilted facets are the three SS ones.
Along $L=0$, they are still completely separated from each other, 
with rough, rounded orientations in between them, and PT transitions as borders.
For $L<0$, facet ridges appear between the three SS facets.
These facet ridges are capped by FRE points.
The local structure at these FRE points changes with temperature 
in an exotic manner as shown in Fig.~\ref{FREshape}.
The global crystal shape structure is less interesting.
Initially it contains the central flat facet but this shrinks 
and vanishes across the roughening line. 

Fig.~\ref{FREshape}(a) shows the local structure 
around the FRE point in region {\bf II}. 
Two PT lines emerge from the FRE point and
form the borders between the SS facets and the rough phase. 
Moreover, a FOR line emerges from the FRE point 
inside the rough phase. 
At this line the crystal orientation angle jumps.
The skipped unstable surface orientations are associated with local  
ZZ facet type zig-zag step configurations. 
Those are unfavourable for $L<0$ and large $K$.
The FOR line is definitely resolved in our numerical data, 
but  remains extremely short (section 5).  At its
longest extension, it reaches only about two percent of
the distance from the FRE point to the center.    
In some parts of region {\bf II} it is so short that it can only
barely be distinguished numerically, if at all.

The FOR lines disappear at the boundary between regions {\bf II} and {\bf III}.
This boundary looks sharp in Fig.~\ref{phshapes},
but in reality it must be a smooth type of crossover.
We discuss this in section 6.

In region {\bf III}, see Fig.~\ref{phshapes}, the FOR line vanished.
The border between the SS facets and the rough phase changed as well.
Close to the FRE point the SS-rough boundaries  are now first-order.
They continue as PT lines beyond PTE-type critical endpoints.
Fig.~\ref{FREshape}(b) shows this local structure.

The first-order segments become longer towards region {\bf IV}.
At the {\bf III}-{\bf IV} phase boundary, two PTE points from opposite
FRE sides meet, and in region {\bf IV}, the entire SS-rough 
phase boundary is first-order.
In the mean time, the central rough region  has been shrinking, 
and it disappears completely at the {\bf IV}-{\bf V} phase boundary.

\section{Numerical Methods}\label{sec:4}

We determine the crystal shape by means of the transfer matrix method.
This means that we calculate the exact free energies 
for semi-infinite lattices 
in terms of 
the largest and nearby eigenvalues of the transfer matrix.
This is a standard method. Therefore we only need to comment here 
on the details concerning the surface tilt fields $E_i$, the amount of 
surface tilts $Q_i$ they create, and the Wulff construction
(Legendre transformation) between these two types of quantities.

There are two natural choices for the directions 
in which the lattice is infinite.
$T$ can be aligned with the $y$-axis or the $x$-axis,
see Fig.~\ref{lattice}. 
We refer to these as $T_x$ and $T_y$ type transfer matrices. 
In $T_x$, the infinite lattice direction coincides with the
$y$-axis and the finite direction with the $x$-axis.
In $T_y$ it is the other way around.
Notice that for each choice there are three rotationally equivalent ones.
For $T_x$, we are able to find the largest eigenvalues for $4 \leq N \leq 18$,
where $N$ is the number of vertices in the time slice.  For $T_y$, we
can calculate the largest eigenvalues for $2 \leq N \leq 10$.

Consider the free energy as function of tilt angles $f(Q_x, Q_y)$. 
This is the  free energy of a facet at a given orientation $(Q_x, Q_y)$. 
Some orientation ranges are thermodynamically unstable. 
These angles are not represented in the crystal shape.
They are skipped, and result in sharp edges in the surface.
To obtain the crystal shape one needs to minimize $f(Q_x, Q_y)$
under the constraint that the volume of the crystal 
(the amount of matter) is conserved.
This leads to the famous Wulff construction~\cite{Wulff}
which  is in essence a geometric construction for the 
Legendre transformation of $f(Q_x, Q_y)$.
The tilt fields $E_i$ are the conjugate variables to the tilt angles $Q_i$.
The crystal shape is a direct representation 
of the free energy function $f(E_x, E_y)$ at a certain temperature 
(at specific values of $K$ and $L$)~\cite{Wortis}.  
The shape functions $Q_x(E_x,E_y)$ and $Q_y(E_x,E_y)$ 
tell us which surface orientations are represented in the crystal.

The transfer matrix provides us with a mixed version of this.
It leads to $f(Q_\perp, E_\parallel)$ and  $Q_\parallel(Q_\perp, E_\parallel)$.
We control the tilt angle in the finite lattice direction, $Q_\perp$,
and the tilt field in the infinite direction, $E_\parallel$.
$Q_\parallel$ sets itself.
$Q_\perp$ is set by the boundary condition in the finite lattice direction,
$h(x_\parallel, x_\perp+N)=h(x_\parallel,x_\perp)+a $.
The total tilt of the surface can take the values 
$a= 0,\pm 2,\dots$ up to $N$ 
when the transfer matrix is aligned with the $y$-axis.
This means that an average slope $Q_x = \sqrt{3} a / 2N$
is  frozen into the surface in the $x$ direction.
In the opposite set-up, with $T_y$, where the lattice is infinite 
in the $x$ direction, a tilt  $Q_y = a/2N$
is  frozen into the surface in the $y$-direction.

\begin{figure}
\centerline{\epsfxsize=8cm \epsfbox{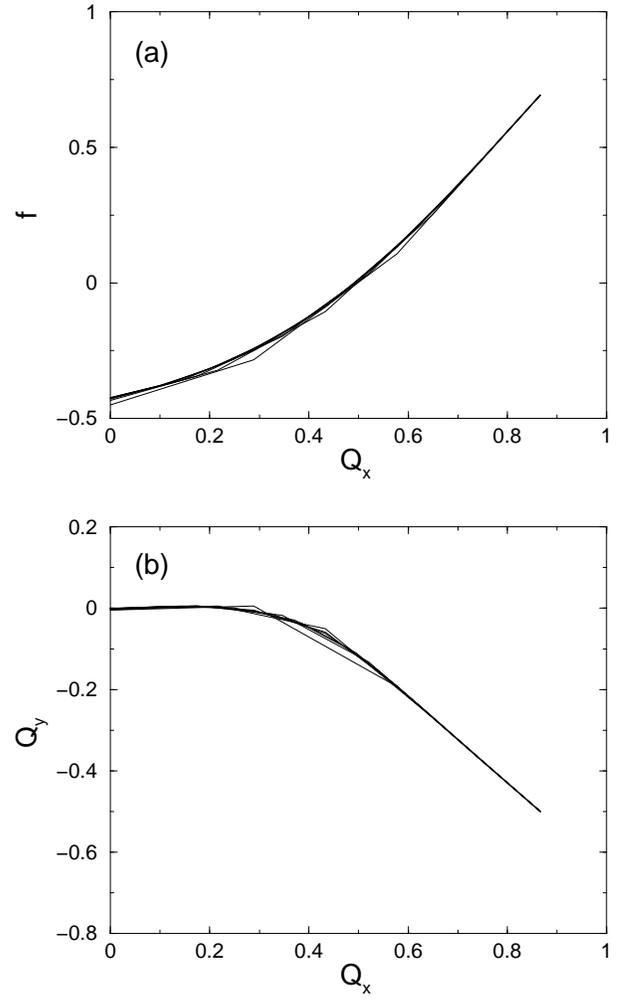}}
\caption{ (a) Free energy $f(Q_x)$ and (b) the vertical surface tilt
$Q_y(Q_x)$ as function of horizontal tilt $Q_x$
at point $K=-0.24$, $L=0.29$  in region {\bf I}
for $E_y=0$ using the $T_x$ transfer matrix set-up
where $Q_x$  takes only discrete values.  Data is shown
for system
sizes $ 6\leq N \leq 18 $.}
\label{Tx-I-hor}
\end{figure}

Consider the transfer matrix at zero tilt field $E_\perp=0$ 
and a specific value of $E_\parallel$.
We determine the largest eigenvalue, $\lambda_0$ of $T$  for each value 
of $Q_\perp$. 
The free energy  density  is equal to 
$f(E_\parallel,0, Q_\perp) = -\log(\lambda_0)/N$.
Fig.~\ref{Tx-I-hor}(a) shows a typical example,
at $K=-0.24$, $L=0.29$, and $E_\parallel=0$.
Only the $Q_\perp>0$ side is shown, 
because the curve is mirror symmetric. 
The discrete set of possible tilt angles $Q_\perp$ 
increases with the strip width $N$, and we need to 
perform a finite size scaling (FSS) analysis to obtain the infinite-by-infinite
lattice result.  Fortunately these FSS corrections are reasonably small, 
and typically converge smoothly. 

The tilt angle $Q_\perp$ that is realized 
is the one that minimizes the free energy. 
This corresponds to the lowest point in Fig.~\ref{Tx-I-hor}(a).
There is no need to repeat the calculation for other values $E_\perp \neq 0$.
They are related by
$f(E_\parallel, E_\perp, Q_\perp) = f(E_\parallel,0, Q_\perp)+E_\perp Q_\perp$
because $Q_\perp$ is conserved by the transfer matrix.
In other words, the crystal shape
$f(E_\perp,E_\parallel)$ and the  shape function $Q_\perp(E_\perp,E_\parallel)$ 
follow from the Legendre transform
of $f(E_\parallel, Q_\perp)$ with respect to $Q_\perp$

\begin{figure}
\centerline{\epsfxsize=8cm \epsfbox{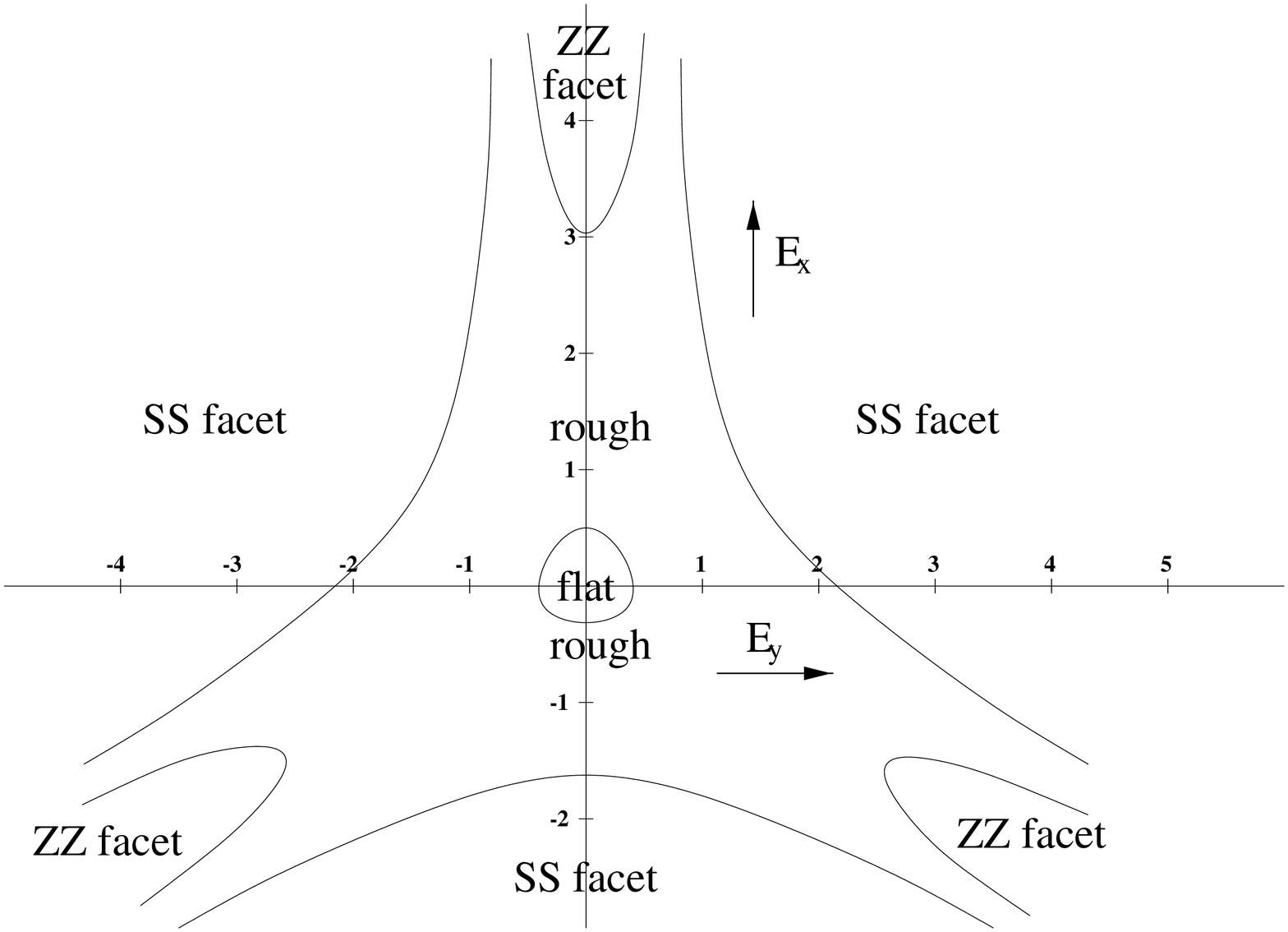}}
\caption{ Global crystal shape at point
$K=-0.24$ and $L=0.29$ in region {\bf I}.}
\label{shape1}
\end{figure}

It is easy to see from graphs like  Fig.~\ref{Tx-I-hor}(a) how 
$Q_\perp$ varies with $E_\perp$. 
Simply rotate the page slowly and watch the
minimum shift, and sometimes skip certain orientations.
In  Fig.~\ref{Tx-I-hor}(a) such skips do not take place.
We follow the $E_x$-axis at $E_\parallel=E_y=0$ in Fig.~\ref{shape1}
along which there are no orientational discontinuities.
Notice  that $f(E_\parallel=0,Q_\perp)$ develops a cusp at 
$Q_x=0$ with system size. 
This means that the central (001) facet is stable at small $E_\perp$. 
The (001)facet boundary is a PT transition because the 
cusp is convex at $Q_\perp=0$  and no orientations are being skipped for 
increasing $E_\perp$ (rotate the page).
The KT roughening transition takes place at values $K$ and $L$ where
the cusp disappears. 
In the rough phase, $f(E_\parallel,Q_\perp)$ has a quadratic minimum at
$Q_\perp=0$. Its curvature represents the roundness of the crystal 
shape at that orientation.

Next, consider the stability of the SS fully tilted facet.
At maximum tilt angle 
the free energy $f(E_\parallel,Q_\perp)$ 
has a definite slope. This means that the SS facet is stable
for a specific range of $E_\perp=E_x$. 
The absence of such a slope would imply that the SS facet 
is thermodynamically unstable, and would be absent in the crystal shape.
The SS-rough phase boundary is a PT transition
because $f(E_\parallel,Q_\perp)$ is convex at $Q_\perp$ near
its maximum 
and no values of $Q_\perp$ are being skipped by the Legendre transformation.

The above analysis is still incomplete because we did not check 
how $Q_\parallel$ is shifting simultaneously.
The function $Q_\parallel(E_\parallel, Q_\perp)$  is contained in the right and 
left eigenvector of the transfer matrix assocated with the largest eigenvalue.
In the interpretation of $T$ as a quantum mechanical 
time evolution operator, $Q_\parallel$ is a velocity type
off-diagonal expectation value.

Fig.~\ref{Tx-I-hor}(b) shows $Q_\parallel(E_\parallel, Q_\perp)
=Q_y(E_y, Q_x)$
at our example point. It shows that the orientation of the 
step excitations  in the surface rotates as expected, 
from vertical just outside the $Q_\perp=Q_\parallel=0$ flat facet
to the correct  
slope just outside the 
SS fully tilted facet.
  
\begin{figure}
\centerline{\epsfxsize=8cm \epsfbox{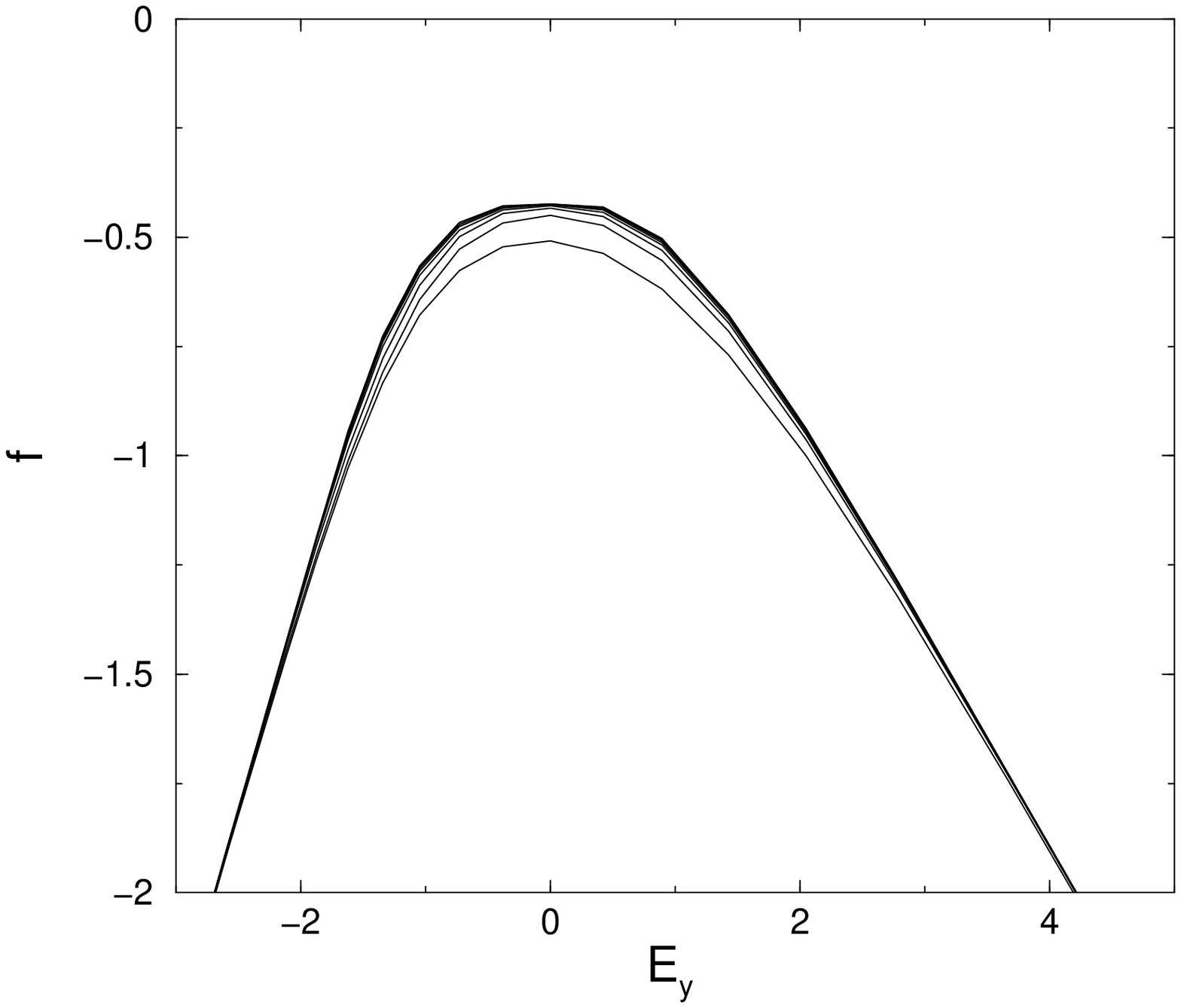}}
\caption{ Free energy $f(E_y)$ as function of the surface tilt field  $E_y$
at point $K=-0.24$, $L=0.29$, in region {\bf I} 
for $Q_x=0$ using the $T_x$ transfer matrix set-up
where $Q_y$ is a continuous variable.  Data is shown
for system
sizes $6 \leq N \leq 18$.}
\label{Tx-I-ver}
\end{figure}

In the above example, we used the $T_x$ transfer matrix set-up 
to investigate the crystal shape in the horizontal direction.
In that set-up $Q_x$ takes only discrete  values, $Q_x = \sqrt{3} a / 2N$.
The opposite set-up with $T_y$ would have been better, since then 
$Q_x$ would have been continuous and $Q_y$ would have been rational.
Fig.~\ref{Tx-I-ver} illustrates this. We still use $T_x$ but now
determine the crystal structure along the $E_y$ axis.
The free energy curve lacks mirror symmetry.
It shows the SS facet at the $E_\parallel<0$ side 
and the ZZ facet at the $E_\parallel>0$ side.  It also shows
the central facet around $E_\parallel=0$.
The facet edges are smooth since they are PT-like.

From the above example it must be clear how
we determined the development of the crystal shape with 
$K$ and $L$, described in section 3.
It required numerous  graphs of the type shown in Fig.~\ref{Tx-I-hor}
and Fig.~\ref{Tx-I-ver}.
The results are quantitively represented in Fig.~\ref{phshapes},
Fig.~\ref{shape1}, and Fig.~\ref{FREshape}.  The latter two 
show the crystal shapes
at actual values of $K$ and $L$ as specified in the figure captions.
In the following two sections we limit the discussion to the most interesting
features  of the crystal shapes:
the FOR lines,
and the PTE points
(see Fig.\ref{FREshape}).

\section{region {\bf II}: The FOR lines}\label{sec:5}

\begin{figure}
\centerline{\epsfxsize=8cm \epsfbox{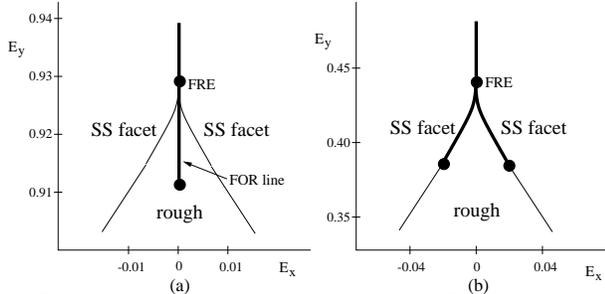}}
\caption{ Local  crystal shape near the FRE point (a) at
point $K=1.43$, $L=-0.63$ in region {\bf II} and (b) at
point $K=1.18$, $L=-0.58$ in region {\bf III}.
}
\label{FREshape}
\end{figure}

In region {\bf II} the crystal shape contains 
only SS-type, fully-tilted facets,
with facet ridges in between them, capped by FRE points.
The only structural difference with the KPZ-type
FRE points in the square lattice
BCSOS model is the presence of the FOR lines sticking into the rough phase, 
compare Fig.~\ref{FREshape}(a)  with Fig.~\ref{6vertex}. 

Such first-order ridges inside the rough phase have
never been realized theoretically or experimentally before.
Unfortunately, they remain very short in our model and are
difficult to resolve numerically.
Fig.~\ref{FREshape}(a) is drawn on scale for point $K=1.43$ and $L=-0.63$
where the FOR line is at its longest, but
remains very  short compared to the distance of the FRE point to the center,
$\Delta E / E \simeq 0.02$.
 
\begin{figure}
\centerline{\epsfxsize=8cm \epsfbox{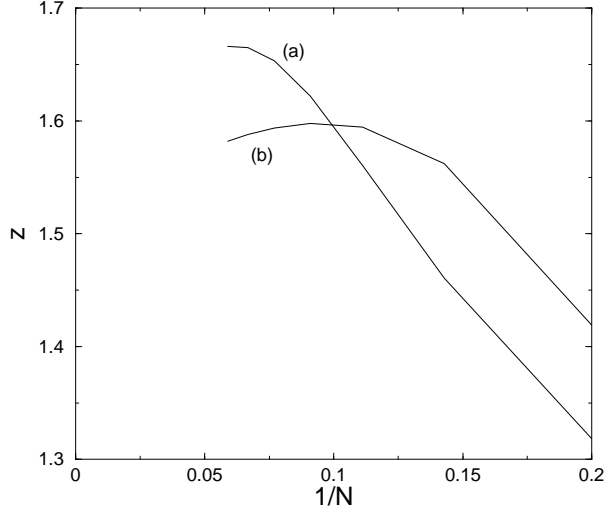}}
\caption{
Finite size scaling behaviour of the mass gap 
scaling exponent $z$, at the endpoint of the FOR line,
at two points in the phase diagram: 
(a) $K=1.43$ and $L=-0.63$ and
(b) $K=0.51$ and $L=-0.42$.
The curves converge to values close 
to the KPZ dynamic exponent $z=\frac{3}{2}$. 
}
\label{FORE-z}
\end{figure}

In the following we provide numerical evidence that FOR lines in our model 
are for real and not a fancy, transient, finite size scaling phenomenon.
This is an important issue.
Suppose the FOR lines vanish in the thermodynamic limit.
Region {\bf II} then represent a realization of the same type of FRE points
as the ones in the square lattice BCSOS model, 
except that the transfer matrix is not stochastic anymore.
We determined the anistropic scaling exponent $z$ from
the leading, $\lambda_0$, and next leading,
$\lambda_1$, eigenvalues of the transfer matrix
in the sector with zero tilt $Q_\perp=0$.
The massgap $m= \log(\lambda_0/\lambda_1)$ 
represents the inverse of the time like correlation length,
$\xi_\parallel \simeq m^{-1} \sim  N^z$.
We do this at our best estimates for the endpoint 
of the FOR line at several values of $K$ and $L$,  and find
$z\simeq 1.6\pm0.1$ (see Fig.~\ref{FORE-z}).
This is close to the KPZ exponent $z=3/2$.
The assumption that the  FOR lines vanish in the thermodynamic limit
would confirm that KPZ scaling is valid beyond the stochastic subspace.
However, in that interpretation, it remains unclear 
how far we are from the stochastic
subspace.  Crossover corrections to scaling might mask
the ``true" non-stochastic exponent $z$.

\begin{figure}
\centerline{\epsfxsize=8cm \epsfbox{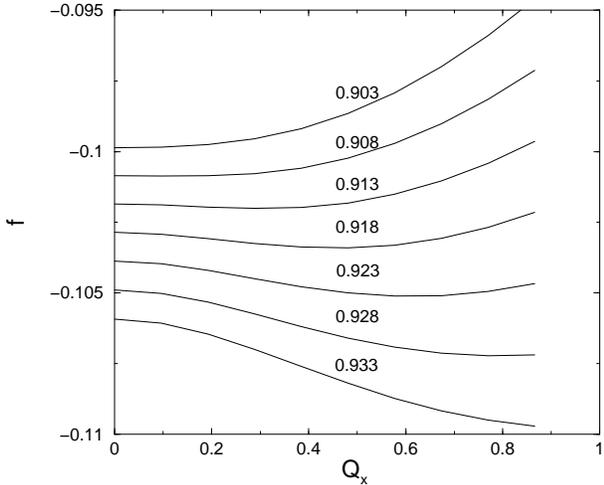}}
\caption{
Free energy as function of the perpendicular tilt $Q_x$
at system size $N=18$ 
along the FOR line at point
$K=1.43$ and $L=-0.63$ in region {\bf II}. The value
of $E_y$ is shown above each curve.
}
\label{W-II}
\end{figure}

The existence of the FOR lines is much more likely, however.
Fig.~\ref{W-II} shows the variation of the free energy
$f(Q_x)$ at $K=1.43$ and $L=-0.63$ for various
$E_y$ close to the FRE point in  Fig.~\ref{FREshape}(a).
We need to use the $T_x$ set-up, 
the one with the infinite lattice direction aligned with the FOR line.
The rational values of $Q_y$ in the $T_y$ set-up skip across the FOR line
and make it invisible altogether (since it is so short). 
In the $T_x$ set-up the $Q_x$ are rational.
In Fig.~\ref{W-II} we show only the data for $N=18$, 
otherwise the graph becomes too crowded.

\begin{figure}
\centerline{\epsfxsize=8cm \epsfbox{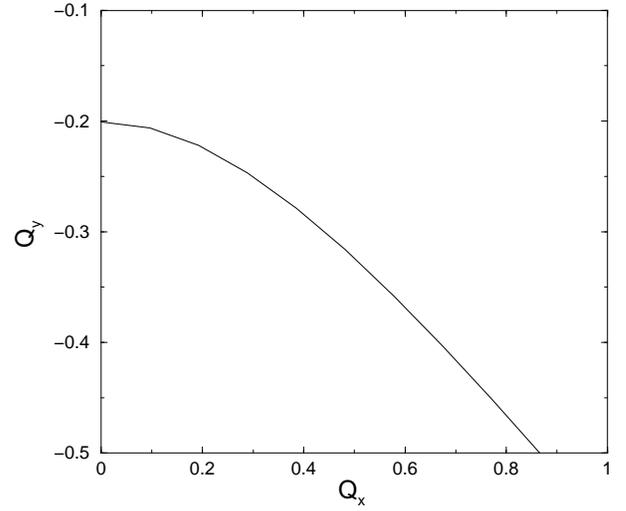}}
\caption{
The tilt angle $Q_y(Q_x)$ at system size $N=18$ 
for $K=1.43$ and $L=-0.63$ in region {\bf II}
when $E_y=0.918$.}
\label{W-II-Qy}
\end{figure}

Fig.\ref{W-II-Qy} shows the corresponding $Q_y(Q_x)$ behaviour at one
value of $E_y$ (the curves for the other values of $E_y$
along the FOR line lie almost on top of this curve).
Well inside the rough phase 
(below the FOR line in Fig.~\ref{FREshape}(a)),
the $f(Q_x)$ curves are convex.
The minimum is at $Q_x=0$, and it shifts smoothly with $E_x$ (rotate the page).
Moreover, the curve remains convex all the way into the wings,
implying PT-type boundaries with the SS facets.

In the opposite limit, well above the FRE point, the curves are concave.
The two  SS facets are coexisting at $E_x=0$, the  entire $f(Q_x)$
curve except for its boundary values in the wings represent thermodynamically 
unstable orientations.

\begin{figure}
\centerline{\epsfxsize=8cm \epsfbox{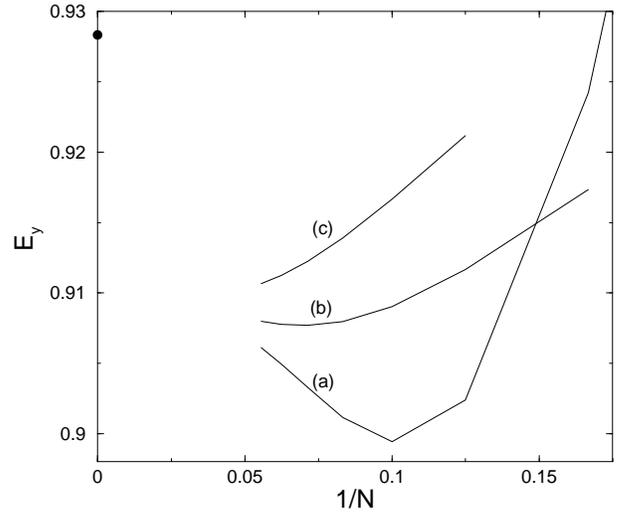}}
\caption{
Finite size scaling approximations for the location of the endpoint of the 
FOR line at $K=1.43$ and $L=-0.63$.  The curves show the value of
$E_y$ where $\lambda_0(Q_x)=\lambda_0(Q_x+\sqrt{3}/2N)$ for:
(a) $Q_x=0$
(b) $Q_x=\sqrt{3}/2N$, and
(c) $Q_x=2 \sqrt{3}/2N$. 
The dot shows the location of the FRE point.
}
\label{FORE}
\end{figure}

The intermediate behaviour of the $f(Q_x)$ curve 
determines the crystal shape close the FRE point.
In the square lattice BCSOS model it flips at once from completely convex 
to completely concave  
(like changing the sign  of $a$ in a polynomial like $f(Q_x)=a Q_x^2$). 
At that FRE point the transfer matrix is stochastic, implying that all 
surface orientations have the same free energy.
In general, $f(Q_x)$ has more structure.
There are many possible structures, but we observe in our 
model the simplest ones.
In Fig.~\ref{W-II}, the curvature of the central part of
the $f(Q_x)$ curve changes before its SS facet wings come down.
This creates the $W$-type shapes,
and therefore gives rise to the FOR line.
All of this happens in a very narrow $E_y$ interval.
In Fig.~\ref{FORE} we show the location of $E_y$ where the curvature at
the central part becomes zero, as function 
of $N$ for $K=1.43$ and $L=-0.63$. 
To be more precise, we show the location of 
$\lambda_0(Q_x)=\lambda_0 (Q_x+\sqrt{3}/2N) $ for $Q_x=0$, $Q_x=\sqrt{3}/2N$,
and $Q_x=2 \sqrt{3}/2N$.
The  curves pass the FRE point at small $N$, but then bend backward.
Still, they do not converge to the FRE point 
(its location is known exactly and shown in the figure for reference).
The curve for $Q_x=0$
is straightening out for system sizes $12 \leq N \leq 18$ and 
converges to a point below the
FRE point at $E_y=0.910 \pm 0.005$.
We conclude from this that the FOR lines must be for real.

\section{Region {\bf III}: Scaling at PTE points}\label{sec:6}

\begin{figure}
\centerline{\epsfxsize=8cm \epsfbox{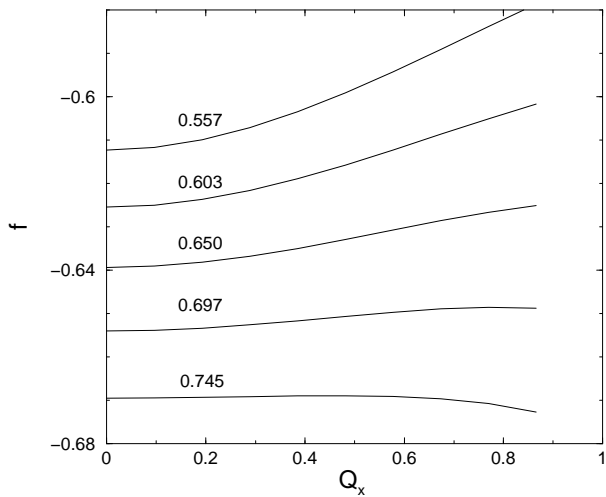}}
\caption{
Free energy as function of the perpendicular tilt $Q_x$
for system size $N=18$
between the FRE and PTE points
at $K=0.37$ and $L=-0.26$ in region {\bf III}.  The value
of $E_y$ is shown above each curve.
}
\label{M-III}
\end{figure}

\begin{figure}
\centerline{\epsfxsize=8cm \epsfbox{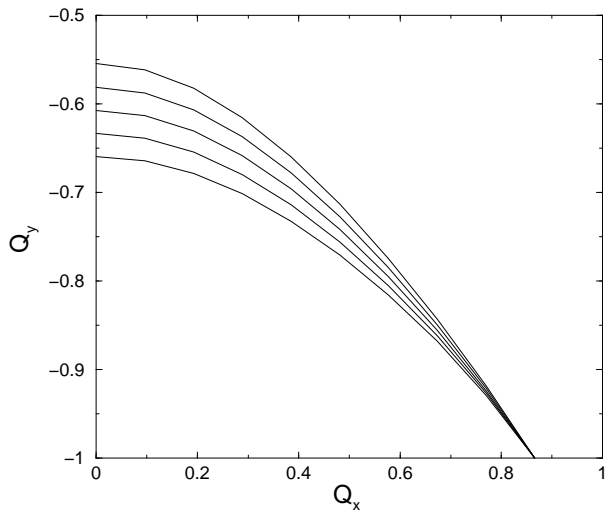}}
\caption{
The tilt angle $Q_y(Q_x)$ at the points in the previous figure.  The
topmost curve corresponds to $E_y=0.557$.
}
\label{M-III-Qy}
\end{figure}

In region {\bf III} the  $f(Q_x)$ curves near  the FRE point
are $M$ shaped (inverted $W$'s).
This is shown in Fig.\ref{M-III} for $K=0.37$ and $L=-0.26$.
Fig.\ref{M-III-Qy} shows the corresponding $Q_y(Q_x)$ plots.
The SS wings of the $f(Q_x)$ curve bend down 
and cross the central minimum before 
the curvature at $Q_x=0$ has a change to change sign.
This behaviour implies that the $f(Q_x)$ curves become concave in the wings,
and therefore that the PT transitions 
become first-order, see Fig.~\ref{FREshape}(b).
Contrast this with the $W$ shapes in region {\bf II}, Fig.~\ref{W-II},  
where the curvature of the central part of $f(Q_x)$ changes sign before 
its wings come down, and thus creates the FOR line.
 
\begin{figure}
\centerline{\epsfxsize=8cm \epsfbox{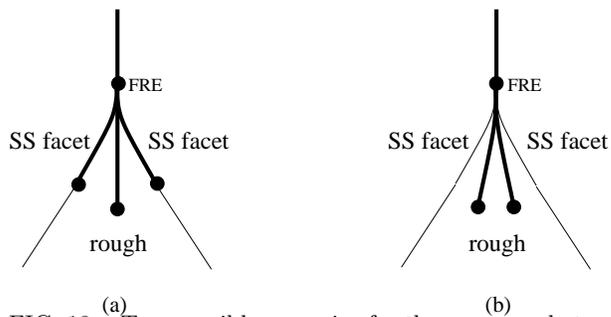}}
\caption{
Two possible  
scenarios for the crossover between region {\bf II} and {\bf III}.
}
\label{split}
\end{figure}

The phase boundary between regions {\bf II} and {\bf III},
can not be sharp. It looks abrupt in Fig.~\ref{phshapes},
but that is an artifact of the shortness of the FOR lines.
This changeover could take many forms. 
The FOR lines require the development of a concave segment in the 
center of $f(Q_x, E_y)$ 
while the PTE points require the development of
concave segments in the wings.
These two phenomena could be independent, and that would result in
both a FOR line and first-order SS-rough segments,
as shown in Fig.\ref{split}(a).
The FRE points become first-order as well. 
This means that subtle developments in the  central part will  be skipped 
and not be expressed in the crystal shape if they take place after the
abrupt crossover from the central minimum to the wing minima.
Our  FOR lines are very short and
move rapidly into the thermodynamically unstable
region to vanish from the crystal shape.
But this is only one of the possible scenarios.

We can not resolve accurately how the 
structure evolves from {\bf II} to {\bf III}.
It happens too fast and the FOR lines are too short. 
We find some evidence that
two concave bubbles move rapidly
along the $f(Q_x, E_y)$ curves from the center to the wings
in opposite directions.
In that case the  FOR line splits into two first-order lines, as
shown in Fig.~\ref{split}.
Each line is still inside the rough phase, and each merges with the PT lines
to form the first-order SS-rough segments.

The scaling behaviour at the PTE critical endpoints in region {\bf III}
deserves attention. 
The existence of such points has been anticipated~\cite{Rottman} but
not realized in solid-on-solid models before, 
and their scaling properties have not been confirmed numerically before.

It is advantageous to align 
the transfer matrix 
along the preferred direction of the steps.
Therefore, we switch our attention to the SS-rough phase boundaries
where, inside the SS facet, the steps  run parallel to the $x$ axis,
see Fig.~\ref{FREshape}(b).
Inside this SS facet, the tilt vector is equal to $\vec Q =(Q_x,Q_y)= (0,1)$
We use the $T_y$ set-up, such that $Q_x$ is continuous.

Consider the crystal shape  close to the PTE point.
Define  $q_y= Q_y^\star- Q_y \geq0$ and $q_x= Q_x- Q_x^\star = Q_x$ 
as the deviation of the tilt with respect to the SS facet 
($Q^\star_y=1$ and $Q^\star_x=0$).
Let  $\epsilon_y=E_y-E_y^\star$ and $\epsilon_x=E_x^\star-E_x$ be
the distances from 
the PTE point
The crystal is rough (faceted) when $\epsilon_y>0$ ($\epsilon_y<0$).
The rough-to-faceted edge is PT-type (first-order) when $\epsilon_x>0$
($\epsilon_y<0$).
The free energy $f(q_x,q_y)$ as function of the tilt angles
is an analytic function near PT transitions. 
This is a direct result of the dilute (free Fermion) nature of the
``gas" of defects~\cite{MdN-IC,Fisher}. 
Therefore we can expand it in terms of a polynomial in $q_x$ and $q_y$.
Our presumption is that this remains true also at the PTE point.
\begin{eqnarray}\label{f-pte}
f(q_x,q_y) = &a_1 |q_x|+ 
  b_1 \epsilon_y |q_y| + \nonumber\\
 & \frac{1}{3} b_3 \epsilon_x |q_y|^3 + 
  \frac{1}{4} b_4 |q_y|^4 +\dots  
\end{eqnarray}
Notice the absence of the quadratic term.
This is a well known property of the entropic hardcore repulsion of 
meandering defect lines (the missing steps in the SS structure)
at low densities~\cite{MdN-IC,Fisher}.
Eq.~\ref{f-pte} reproduces the usual PT scaling behaviour.
The crystal shape follows from minimizing $f(q_x,q_y)$ with respect to 
$q_x$ and $q_y$ at constant $\epsilon_x>0$ and $\epsilon_y>0$.
This gives $q_x=0$ and 
$b_1 \epsilon_y + b_3 \epsilon_x q_y^2 + b_4 q_y^3=0$.
Along a path of constant $\epsilon_x>0$ the tilt angle vanishes
with a square root power, $q_y\sim |\epsilon_y|^{\frac{1}{2}}$ 
and the free energy has the famous PT power law singularity,
$f\sim |\epsilon_y|^{\frac{3}{2}}$. This is the power with 
which the surface orientation connects to the SS facet.
The anisotropic scaling exponent $z=2$,  $l_y\sim l_x^z$,
follows from the $f\sim q_y^3$ power 
with which the free energy scales at the PT points. 
This implies that the free energy of  one single defect scales 
per unit length as
$m\sim l_x [f(2 \pi/l_x)-f(0)] \sim l_x^2$ with $l_x$ 
a finite lattice length in the direction perpendicular to the defect. 
$m$ is inversely proportional to the  correlation length in the 
$y$-direction, $m \sim l_y^{-1}$.
Our numerical results are not shown here, since they are in complete agreement with this.

At the PTE point, $\epsilon_x=0$, the same quantities scale according 
to Eq.(\ref{f-pte}) as $q_y\sim |\epsilon_y|^{\frac{1}{3}}$, 
$f\sim q_y^4$, and $z=3$.
Along the first-order line, $\epsilon_x<0$, the jump in 
the orientation angle vanishes linearly, $\Delta q_y \sim |\epsilon_x|$.
$q_x$ develops a jump as well, but with a weaker power,
because the rotation of the  surface orientation inside the rough phase is an
higher order effect.

It is not certain that $q_y^4$ is the lowest surviving 
power in the free energy at the PTE point. 
It could be the $|q_y|^5$  term instead.  
For example, in the conventional free fermion analysis for PT transitions
the free energy is an odd function in $|q_y|$.
In that case the critical exponents change to  
$q_y\sim |\epsilon_y|^{\frac{1}{4}}$, 
$f\sim q_y^5$, and $z=4$.

\begin{figure}
\centerline{\epsfxsize=8cm \epsfbox{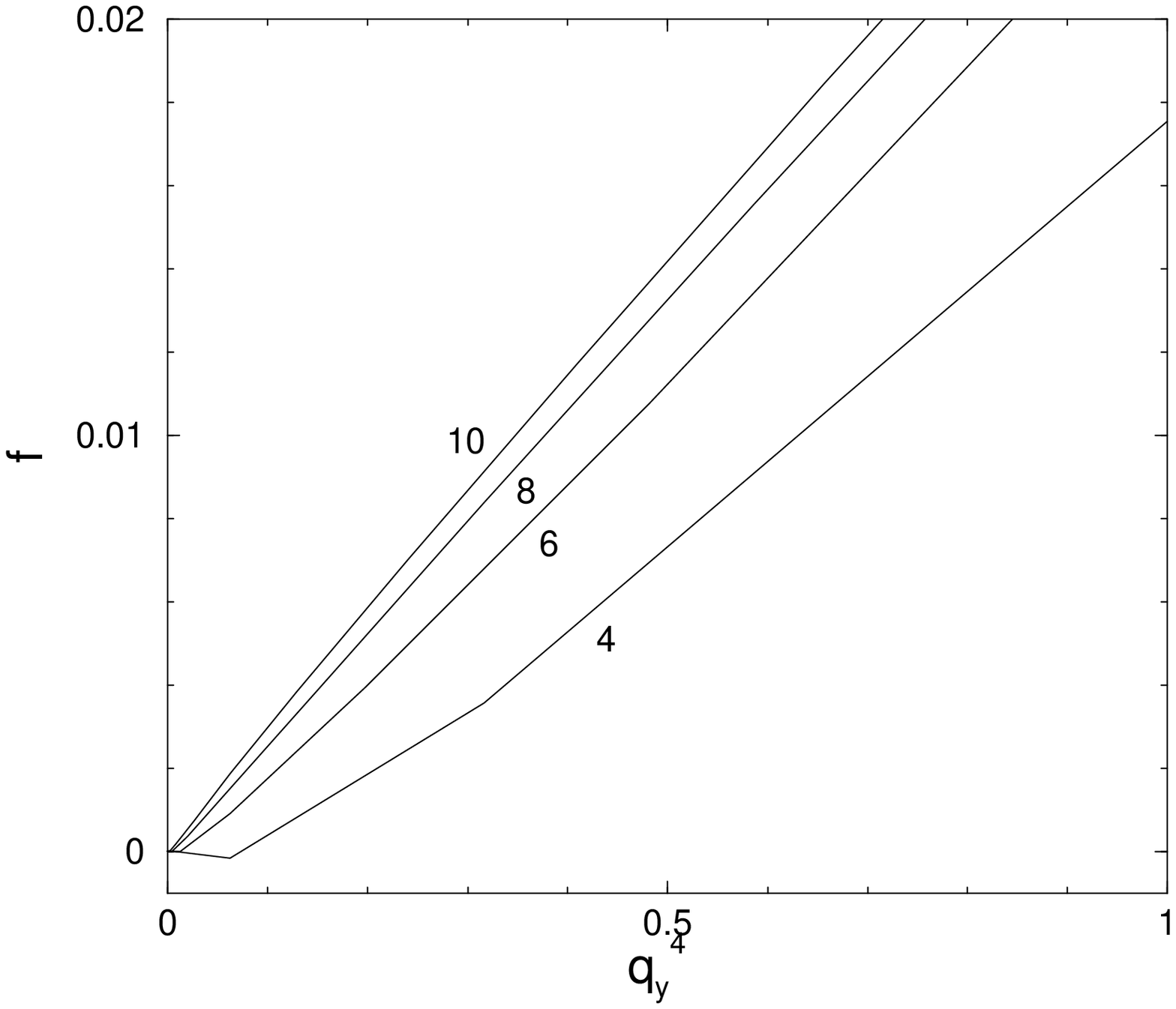}}
\caption{
Free energy as a function of $q^4$ at PTE point $K=0.37$ and $L=-0.26$
as measured by $T_y$ for system sizes $4 \leq N \leq 10$.}
\label{PTE-q4}
\end{figure}

In Fig.~\ref{PTE-q4} we show the free energy as a function
of $q_y^4$ at the PTE point for $K=0.37$, $L=-0.26$ in region
{\bf III} for system sizes $4 \leq N \leq 10$.  
The curves are straight lines.
This confirms that the free energy is going to zero as $f\sim q_y^4$ 
and therefore that  $z=3$.

The scaling behaviour $q_y\sim |\epsilon_y|^{\frac{1}{3}}$  across the 
PTE point and the power $\Delta q_y \sim |\epsilon_x|$
with which the jump in $q_y$ along the first-order boundary
vanishes are difficult to resolve numerically
because of the discrete nature of $q_y$.
We find evidence that the jump scales linearly, but the results
are somewhat ambiguous.

At the {\bf III} to {\bf IV} phase boundary the PT segment
along the rough-SS facet boundary shrinks to zero and the two PTE points merge.
This can be described by a polynomial of the form
\begin{eqnarray}\label{f-2pte}
f(q_x,q_y) =& a_1 |q_x|+ 
  b_1 \epsilon_y |q_y| + \nonumber\\
 & \frac{1}{3} b_3 (c-\epsilon_x^2) |q_y|^3 + 
  \frac{1}{4} b_4 |q_y|^4 +\dots  
\end{eqnarray}
$\epsilon_x$  is now the distance along the rough-SS ridge from the point 
in the middle between the two PTE points, and $c>0$ ($c<0$) in region {\bf III} 
({\bf IV} ). This polynomial, if correct, would imply that the 
critical exponents at the merging point, $c=0$,
are the same as at the PTE points,
$z=3$ and $\beta=\frac{1}{3}$, except that the 
jump in angle vanishes now as, as $q_y \sim |\epsilon_x|^2$.
We find numerically that the  free energy scales as
$f\sim q_y^4$ and, therefore, that indeed $z=3$.

\section{Conclusions}\label{sec:7}

In this paper we obtained the crystal shape for a BCSOS model 
on a honeycomb lattice. This shape is quite complex. 
It contains two types of fully tilted facets, the SS and ZZ facets. 
The latter disappear for negative nearest neighbour interactions.
In region {\bf III} in that part of the phase diagram, we find that  
the rough-to-SS facet boundary becomes partially first-order
with sharp edges where the surface orientation jumps. 
These first-order segments connect
at PTE critical endpoints to conventional PT-type segments
where the  rough phase connects to the facet without a jump in orientation.
The scaling properties  at these PTE points follow from a simple higher-order
polynomial free energy generalization of PT transitions, with anisotropic
scaling with exponent $z=3$ instead of $z=2$.

In region {\bf II} of the phase diagram the rough-SS phase boundary is 
PT-type everywhere, but a first-order line sticks out of the FRE 
point into the rough phase. These FOR lines remain very small in this model, 
but we resolved them numerically sufficiently to be sure they exist.

The main object of this study is to establish how general and universal
the KPZ scaling properties of the FRE point in the square lattice BCSOS 
model are. 
Our conclusion is that the stochastic nature of their transfer matrix 
makes KPZ-type FRE points special and unstable.
In general the local crystal shape around FRE points is more complex
than simply one facet ridge splitting into two PT lines.
Instead there are FOR lines sticking into the rough phase and/or
the  rough-facet phase boundaries become first-order.
Even more complex structures (not observed in our model)  
are possible as well.

We are unable to resolve the scaling  properties of the 
the critical point at the end of the FOR line 
(opposite to the FRE point).  The FOR lines remain too short.
It might still be KPZ-type scaling with $z=\frac{3}{2}$.
In the exactly soluble
square lattice BCSOS model, the KPZ nature of the scaling
is closely linked to the degeneracy of the free energy
with respect to all surface
orientations (this is implied by the
stochasticity of the transfer matrix).  Our model lacks
this special symmetry, but, at these critical points at the
end of the FOR
line, the free energies of small angle surface orientations become
degenerate.  Maybe this local version of that symmetry
is sufficient for $z=\frac{3}{2}$ anisotropic scaling, or it
could lead to a new, yet unknown value of $z$.
This aspect needs further study.
 
This work is supported by NSF grant DMR-9700430.

\end{multicols}

\begin{thebibliography}{30}

\bibitem{Wulff} 
G.~Wulff, Z.~Kristallogr.~Mineral.~{\bf34}, 449 (1901).

\bibitem{Herring} 
C.~Herring, Phys.~Rev.~{\bf 82} 87 (1951).

\bibitem{Wortis} 
For a review see, 
C.~Rottman and M.~Wortis, Phys. Rep. {\bf 103}, 59 (1984).

\bibitem{Rottman}
C.~Rottman and M.~Wortis, Phys.Rev.B {\bf 29}, 328 (1984).

\bibitem{Nolden} 
I.M.~Nolden and H.~van Beijeren, Phys.Rev.B {\bf 49}, 17224 (1994);
I.M.~Nolden, J.Stat.Phys. {\bf 67}, 155 (1992).

\bibitem{Grimb}
R.F.P.~Grimbergen, H.~Meekes, P.~Bennema, H.J.F.~Knops, and M.~den Nijs,
Phys.Rev.~B {\bf 58}, 5258 (1998).

\bibitem{JN-MdN} 
J.~Neergaard and M.~den Nijs, Phys.Rev.Lett. {\bf 74}, 730 (1995).

\bibitem{Buk-Shore}
J.D.~Shore and D.J.~Bukman, Phys.Rev.Lett. {\bf 72}, 604 (1994).

\bibitem{Buk-Shore2}
J.D.~Shore and D.J.~Bukman, Phys.Rev.E {\bf 51}, 4196 (1995).

\bibitem{KT}
See e.g., M.~Kosterlitz and D.J.~Thouless, J.~Phys.~C {\bf 6}, 1181 (1973).

\bibitem{Pok-Tal}
V.L.~Pokrovsky and A.L.~Talapov, Phys.Rev.Lett. {\bf 42}, 65 (1979);
V.L.~Pokrovsky and A.L.~Talapov, Zh.Exp.Teor.Fys. {\bf 78}, 269 (1980).

\bibitem{MdN-IC}
For a review on the Fermion approach see:
M.~den Nijs, in
{\em Phase Transitions  and Critical Phenomena}, Vol.~12,
eds.~ C. Domb and J.L.~Lebowitz (Academic Press, London, 1988).

\bibitem{Fisher}
M.E.~Fisher, Phys.Rev.B {\bf 36}, 644 (1987).

\bibitem{Dhar}
D.~Dhar, Phase Transitions {\bf 9}, 51 (1987).

\bibitem{Gwa-Spohn}
LH.~Gwa and H.~Spohn, Phys.~Rev.~A {\bf 46}, 844 (1992).

\bibitem{DD-MdN}
Douglas Davidson and Marcel den Nijs, in preparation.

\bibitem{MdN-87} 
For more detail, see e.g., M.~den Nijs,
J.~Phys.~A  (1987).


\end{thebibliography}
\end{document}